\documentclass[
twocolumn, 
nofootinbib,
secnumarabic,amssymb, preprintnumbers, superscriptaddress, aps, prd]{revtex4-1}
\usepackage{amsmath,times,amssymb,latexsym,graphics,braket,color}
\newcommand{\ex}[1]{\mathrm{e}^{#1}}

\newcommand{\pa}[1]{\left(#1 \right)}

\newcommand{\BR}[1]{\Biggl[#1 \Biggr]}

\newcommand{\bb}[1]{\mathbb{#1}}

\newcommand{\ca}[1]{\mathcal{#1}}

\newcommand{\dg}[1]{#1^{\dagger}}
\newcommand{\fr}{\frac}
\newcommand{\s}[1]{\sqrt{#1}}

\def\be{\begin{equation}}
\def\ee{\end{equation}}
\def\ba{\begin{eqnarray}}
\def\ea{\end{eqnarray}}

 \def\ep{{\epsilon}}

 \def\ba{{\bar{\alpha}}}

 \def\D{{\Delta}}
 \def\g{{\gamma}}

 \def\e{{\epsilon}}

\def\tr{{\text{tr}}}

\begin{document}
\preprint{YITP-19-60, arXiv:1907.06646}
\title{Dynamics of Entanglement Wedge Cross Section from Conformal Field Theories}
\date{\today}
\author{Yuya Kusuki}\email[]{yuya.kusuki@yukawa.kyoto-u.ac.jp}
\author{Kotaro Tamaoka}\email[]{kotaro.tamaoka@yukawa.kyoto-u.ac.jp}
\affiliation{\it Center for Gravitational Physics, 
Yukawa Institute for Theoretical Physics (YITP), Kyoto University, 
Kitashirakawa Oiwakecho, Sakyo-ku, Kyoto 606-8502, Japan.}

\begin{abstract}
We derive dynamics of the entanglement wedge cross section directly from the two-dimensional holographic CFTs with a local operator quench. This derivation is based on the reflected entropy, a correlation measure for mixed states. We further compare these results with the mutual information and ones for integrable systems. This comparison directly suggests the classical correlation plays an important role in chaotic systems, unlike integrable ones. Besides a local operator quench, we study the reflected entropy in heavy primary states and find a breaking of the subsystem ETH. We checked the above results also hold for the odd entanglement entropy, which is another measure for mixed states related to the entanglement wedge cross section. 
\end{abstract}
\maketitle
\noindent
\section{Introduction and Summary}
The strongly coupled many-body systems, which are typically chaotic, attract a lot of attention in the physics community. One useful tool to capture the dynamics and the thermalization in such systems is the entanglement entropy (EE), which is defined by
\begin{equation}
S(A)=-\tr \rho_A \log \rho_A,
\end{equation}
where $\rho_A$ is a reduced density matrix for a subsystem $A$, obtained by tracing out its complement $A^c$. 
This quantity measures entanglement between subsystem $A$ and its complement $A^c$ if a pure state describes the entire system.
The EE also plays a significant role in quantum gravity via the AdS/CFT correspondence\cite{Maldacena:1997re,Ryu2006a, Ryu2006, Hubeny2007}. Note that the systems with gravity dual (called holographic CFTs) are sometimes referred as ``the most chaotic system''\cite{Maldacena2016} . 

If one focuses mixed states $\rho_{AB}$ associated with a subsystem $AB\equiv A\cup B$ and wishes to measure the correlation between $A$ and $B$, however, we have many measures in the literature and no unique choice as opposed to the EE for pure states. 
Therefore, from both conceptual and practical viewpoints, we should use the one(s) which have a clear meaning in the setup under consideration. 

In this letter, we will focus on the reflected entropy (RE) $S_R$\cite{Dutta2019} which has a sharp (conjectured) interpretation in the context of AdS/CFT. We expect that
\be
S_R(A:B)=2E_W(A:B)
\ee
where $E_W$ is area of the minimal cross section of the entanglement wedge\cite{Takayanagi2018a, Nguyen2018} dual to the reduced density matrix\cite{Czech:2012bh,Wall:2012uf,Headrick:2014cta}. (See also \cite{Bao2018, Umemoto2018a, Hirai2018, Bao2019c, Espindola2018, Bao2018b, Guo2019a, Bao2019b, Yang2019, Kudler-Flam2019, BabaeiVelni2019, Caputa2019a, Tamaoka2019, Guo2019, Bao2019,Liu:2019qje,Du:2019emy,Harper:2019lff,Kudler-Flam:2019wtv} for further developments in this direction.) We will give the definition of the RE in the next section. This bulk object, called entanglement wedge cross section (EWCS), is a natural generalization of the minimal surfaces. In particular, if $B=A^c$ and $\rho_{AB}$ is a pure state, $E_W(A:B)$ reduces to the area of the minimal surfaces associated with the $S(A)(=S(A^c))$. In the same way, $S_R(A:B)$ reduces to the $2S(A)$ for pure states. 

The point is that the RE is expected to be more sensitive to classical correlations than the mutual information $I(A:B)=S(A)+S(B)-S(AB)$, therefore, the RE would be a refined tool to investigate the chaoticity in the light of classical correlations. Thus, it naturally motives us to study entanglement in non-equilibrium situations by both the RE and the EE (the mutual information), and to compare these two measures. This is one of the main interest in the present letter.

Let us summarize the results of the present letter. First, we have studied the time evolution of the RE by a local operator quench and see a perfect agreement with the EWCS for a falling particle geometry\cite{Nozaki2013}. 
Comparing with the mutual information and results for rational conformal field theories (RCFTs), our results directly show that in the dynamical process for chaotic systems, classical correlations play an important role, unlike integrable systems.
From this observation, we can conclude that the comparison between the RE and the mutual information allows us to provide more information about chaotic nature of a given theory than mutual information itself.
Second, our analysis clarifies the bulk dual of the heavy primary state. Remarkably, we find that in holographic CFTs, nevertheless very chaotic systems, such state does not satisfy the subsystem eigenstate thermalization hypothesis\cite{Garrison:2015lva,Dymarsky:2016ntg}. 

We have to mention that the above analysis also holds for the odd entanglement entropy\cite{Tamaoka2019}, which is another generalization of the EE for mixed states. These results can be achieved by using the fusion kernel approach in two-dimensional CFT\cite{Kusuki2018c, Collier2018, Kusuki2019}. We will report the detail of technical parts (for both CFT and gravity) in our upcoming paper\cite{Kusuki:2019e}. 

\section{Reflected entropy}
Here we review the definition of the reflected entropy (RE). We consider the following mixed state,
\begin{equation}
\rho_{AB} = \sum_ {n}   p_n \rho^{(n)}_{AB},
\end{equation}
where each $\rho^{(n)}_{AB}$ represents a pure state as
\begin{equation}
\rho^{(n)}_{AB}=\sum_{i,j} \s{\lambda^i_n \lambda^j_n} \ket{i_n}_A  \ket{i_n}_B  \bra{j_n}_A  \bra{j_n}_B  ,
\end{equation}
where $\ket{i_n}_A \in \ca{H}_A$, $\ket{i_n}_B \in \ca{H}_B$ and $\lambda_n^i$ is a positive number such that $\sum_i \lambda_n^i=1$.
The real number $p_n$ is the corresponding probability associated with its appearance in the ensemble.
For this mixed state, we can provide the simplest purification as
\begin{equation}
\ket{\s{\rho_{AB}} }= \sum_{i,j,n} \s{p_n  \lambda^i_n \lambda^j_n}  \ket{i_n}_A  \ket{i_n}_B  \ket{j_n}_{A^*}  \ket{j_n}_{B^*}   ,
\end{equation}
where $\ket{i_n}_{A^*} \in \ca{H}^*_A$ and $\ket{i_n}_{B^*} \in \ca{H}^*_B$ are just copies of $\ca{H}_A$ and $\ca{H}_B$.  Then, the RE is defined by
\begin{equation}
S_R(A:B) \equiv -\tr \rho_{AA^*} \log \rho_{AA^*},
\end{equation}
where $\rho_{AA^*}$ is the reduced density matrix of $\rho_{AA^*BB^*}=\ket{\s{\rho_{AB}}} \bra{\s{\rho_{AB}}}$  after tracing over $\ca{H}_B \otimes \ca{H}^*_B$.

\section{Setup}
Our interest in this letter is to study a local operator quench state \cite{Nozaki2014,Nozaki2014a}, which is created by acting a local operator $O(x)$ on the vacuum in a given CFT at $t=0$,
\be\label{eq:defO}
\ket{\Psi(t)}=\s{\ca{N}}\ex{-\ep H-iHt} O(x)\ket{0}, 
\ee
where $x$ represents the position of insertion of the operator, $\e$ is a UV regularization of
the local operator and $\ca{N}$ is a normalization factor so that $\braket{\Psi(t)|\Psi(t)}=1$.

\begin{figure}[t]
 \begin{center}
 \resizebox{85mm}{!}{
  \includegraphics{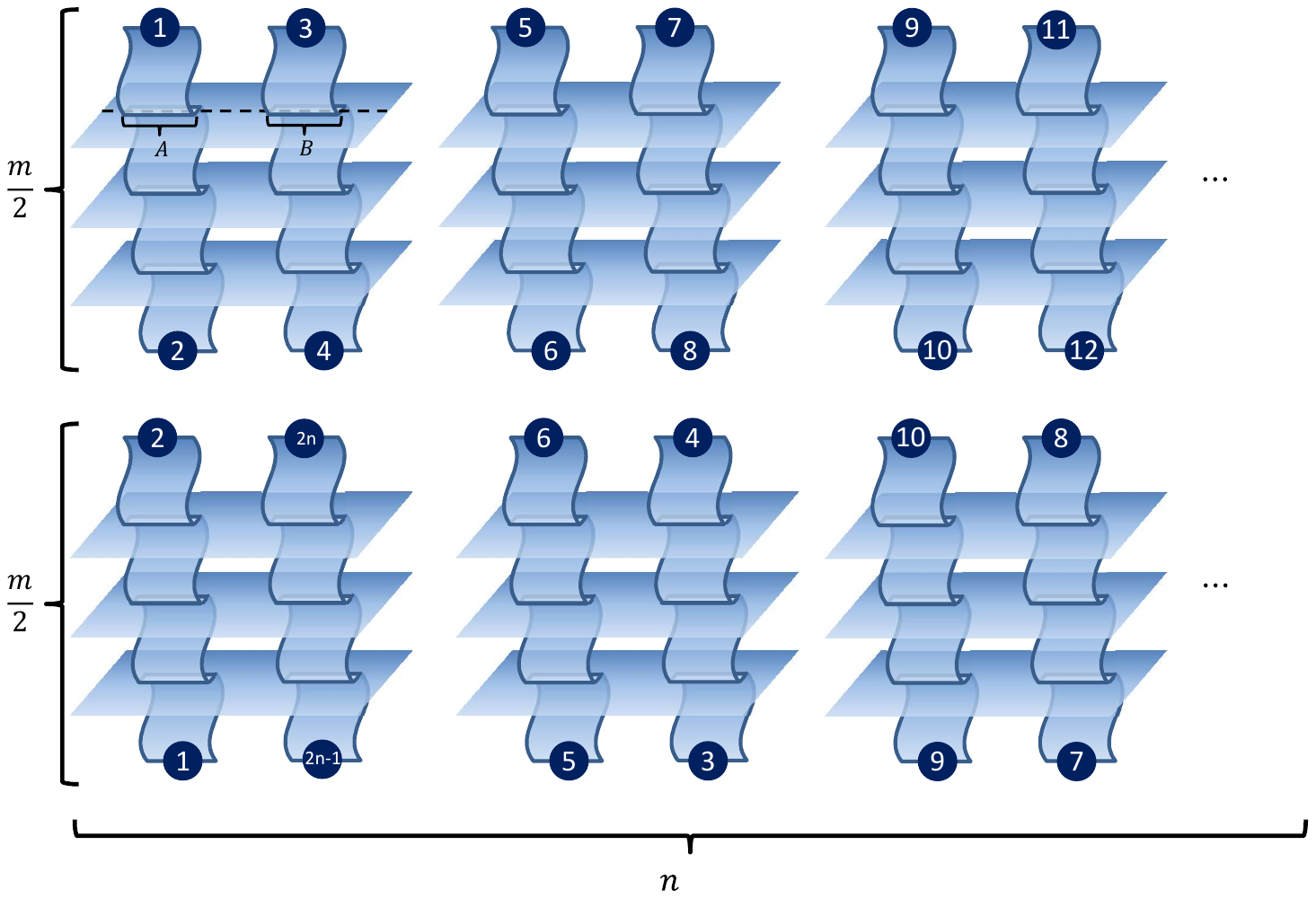}}
 \resizebox{40mm}{!}{ \includegraphics{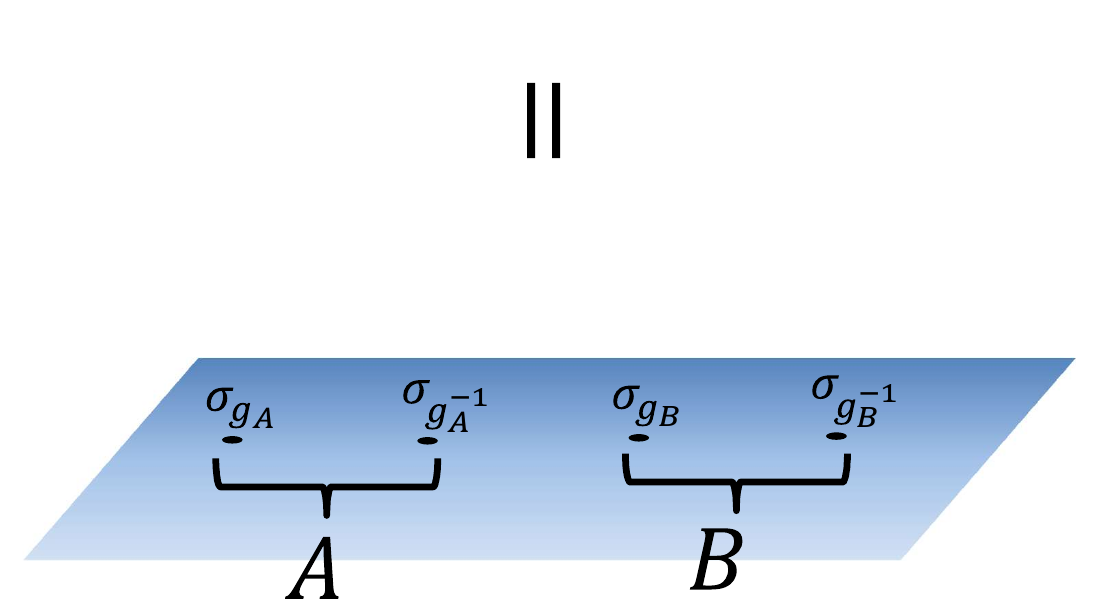}}
  
 \end{center}
 \caption{The path integral representation of the Renyi reflected entropy (for vacuum). Edges labeled with the same number get glued together.
We can instead view it as a correlator with four twist operators $\Braket{\sigma_{g_A}(u_1)\sigma_{g_A^{-1}}(v_1)   \sigma_{g_B}(u_2) \sigma_{g_B^{-1}}(v_2) }_{\text{CFT}^{\otimes mn}}$. 
 }
 \label{fig:replica}
\end{figure}

The RE can be evaluated in the path integral formalism \cite{Dutta2019}.
For example, the Renyi RE in the vacuum can be computed by a path integral on $m \times n$ copies as shown in FIG. \ref{fig:replica}. Here, we would view this manifold as a correlator with twist operators as in the lower of FIG. \ref{fig:replica}, where we define the twist operators $\sigma_{g_A}$ and $\sigma_{g_B}$. 
Here, we focus on the following mixed state,
\begin{equation}
\rho_{AB}=\tr_{(AB)^c} \ket{\Psi(t)} \bra{\Psi(t)},
\end{equation}
where $\Psi(t)$ is a time-dependent pure state as $\ket{\Psi(t)}=\s{\ca{N}}\ex{-\ep H-iHt} O(0)\ket{0} $.
Then, in a similar manner to the method in \cite{Nozaki2014}, the replica partition function in this state can be obtained by a correlator as
\begin{align}\label{eq:Renyi}
&\frac{1}{1-n}\log\frac{Z_{n,m}}{Z_{1,m}^n},
\end{align}
where
\begin{align}
&Z_{n,m}=\Big\langle\sigma_{g_A}(u_1)\sigma_{g_A^{-1}}(v_1)  {O^{\otimes mn}}(w_1,\bar{w}_1)\nonumber\\&\hspace{2cm}\dg{{O^{\otimes mn}}}(w_2,\bar{w}_2)\sigma_{g_B}(u_2) \sigma_{g_B^{-1}}(v_2) \Big\rangle_{\text{CFT}^{\otimes mn}}.
\end{align}
Here we abbreviate $V(z,\bar{z})\equiv V(z)$ if $z\in\bb{R}$ and the operators $O$ are inserted at
\begin{equation}
w_1=t+ i \e, \ \ \ 
\bar{w}_1=-t+ i \e, \ \ \ 
w_2=t- i \e, \ \ \ 
\bar{w}_2=-t- i \e. \ \ \ 
\end{equation}
To avoid unnecessary technicalities, we do not show the precise definition of the twist operators $\sigma_{g_A}$ and $\sigma_{g_B}$ (which can be found in \cite{Dutta2019}) because in this letter, we only use the scaling dimension of the twist operators,
\begin{equation}
\begin{aligned}
&h_{\sigma_{g_A}}=h_{\sigma_{g_A^{-1}}}=h_{\sigma_{g_B}}=h_{\sigma_{g_B^{-1}}}=\fr{cn}{24} \pa{m-\fr{1}{m}}  (= n h_m)   ,\\
&h_{\sigma_{g_A^{-1} g_B}} = h_{\sigma_{g_B^{-1} g_A}}= \fr{c}{12}\pa{n-\fr{1}{n}} (= 2h_n)  .
\end{aligned}
\end{equation}
Here $O^{\otimes N}\equiv O\otimes O\otimes\cdots \otimes O$ is an abbreviation of the operator on $N$ copies of CFT ($\text{CFT}^{\otimes N}$). 
We will take $n,m\rightarrow1$ limit so that the \eqref{eq:Renyi} reduces to the original RE.

\section{Holographic CFT}
\begin{figure}[t]
 \begin{center}
  \resizebox{80mm}{!}{\includegraphics{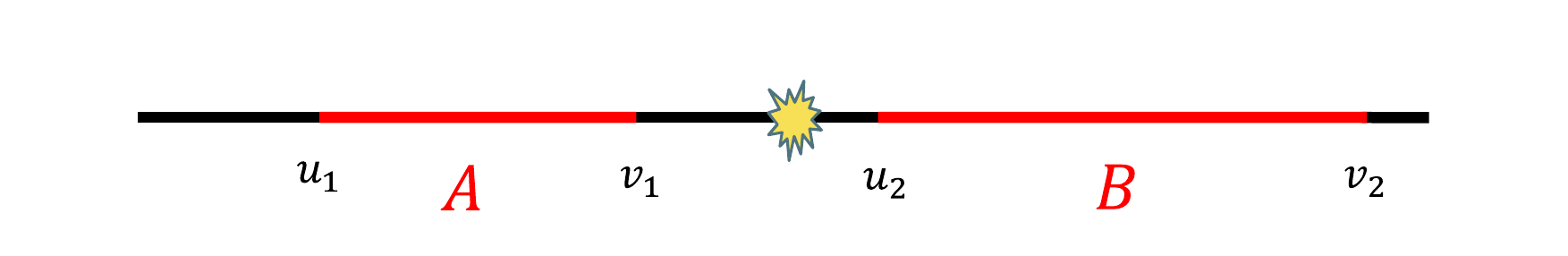}}
 \end{center}
 \caption{We study the setup $0< u_2<-v_1<-u_1<v_2$. We excite the vacuum by acting an local operator on $x=0$ at $t=0$.}
 \label{fig:setup}
\end{figure}
As a concrete example, we consider the setup described in FIG. \ref{fig:setup}. Namely, we set our subregion $A=[u_1,v_1], B=[u_2,v_2]$ and assume $0<\e\ll u_2<-v_1<-u_1<v_2$.

\begin{figure}[t]
 \begin{center}
 \resizebox{85mm}{!}{
 \includegraphics{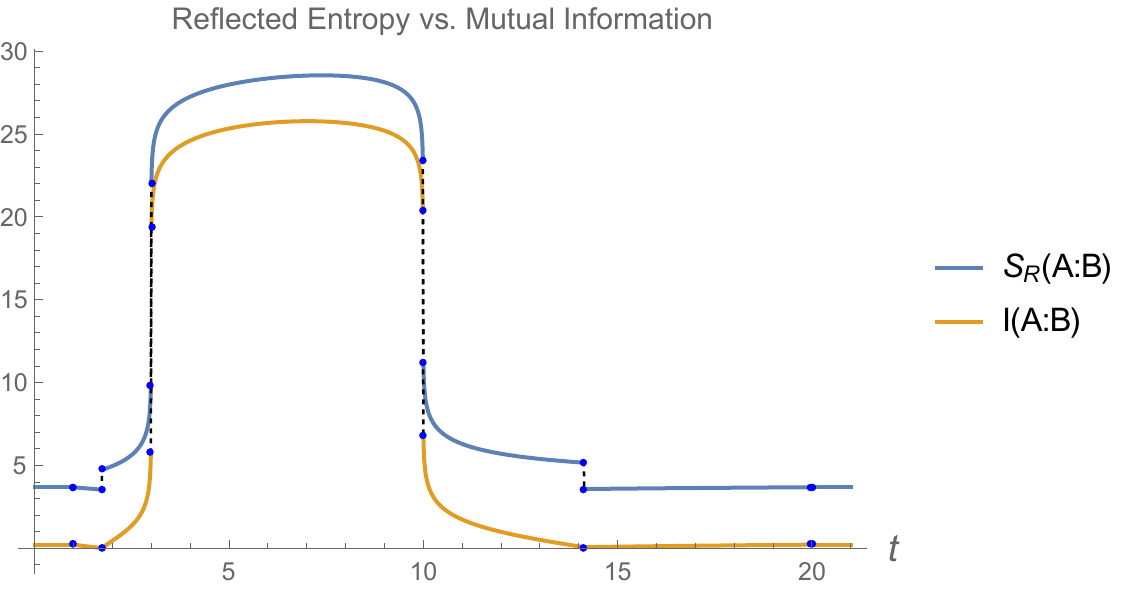}
 \includegraphics{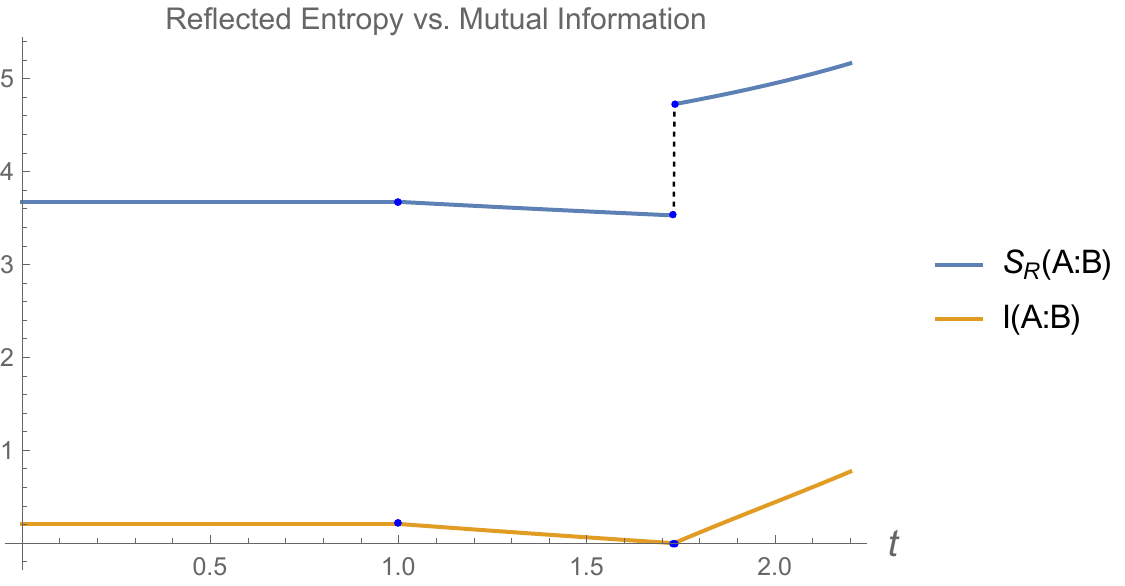}}
 \vspace{5mm}\\
 \resizebox{85mm}{!}{
  \includegraphics{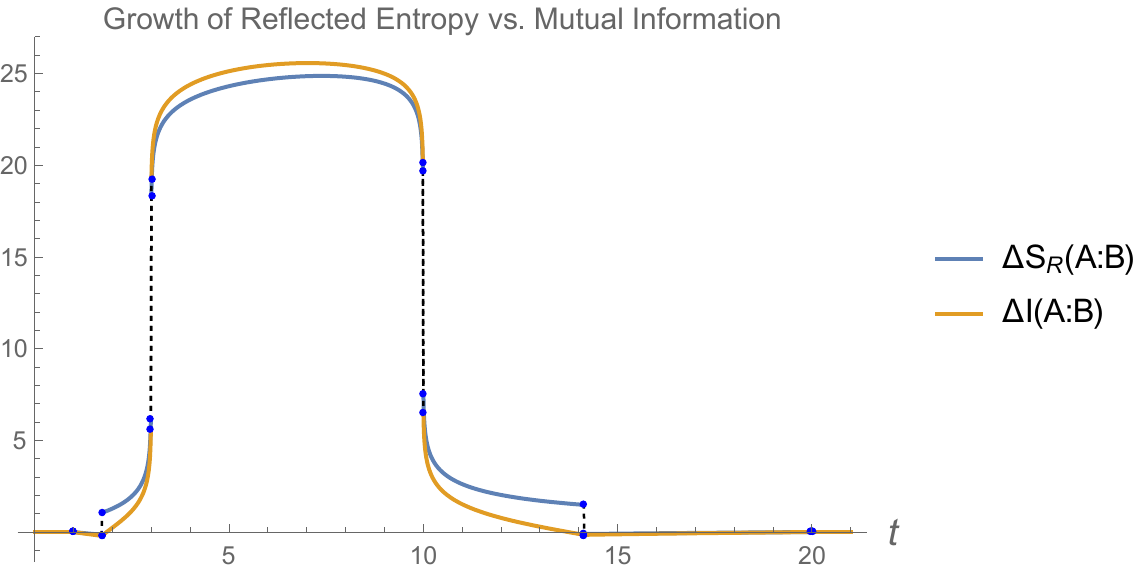}
  \includegraphics{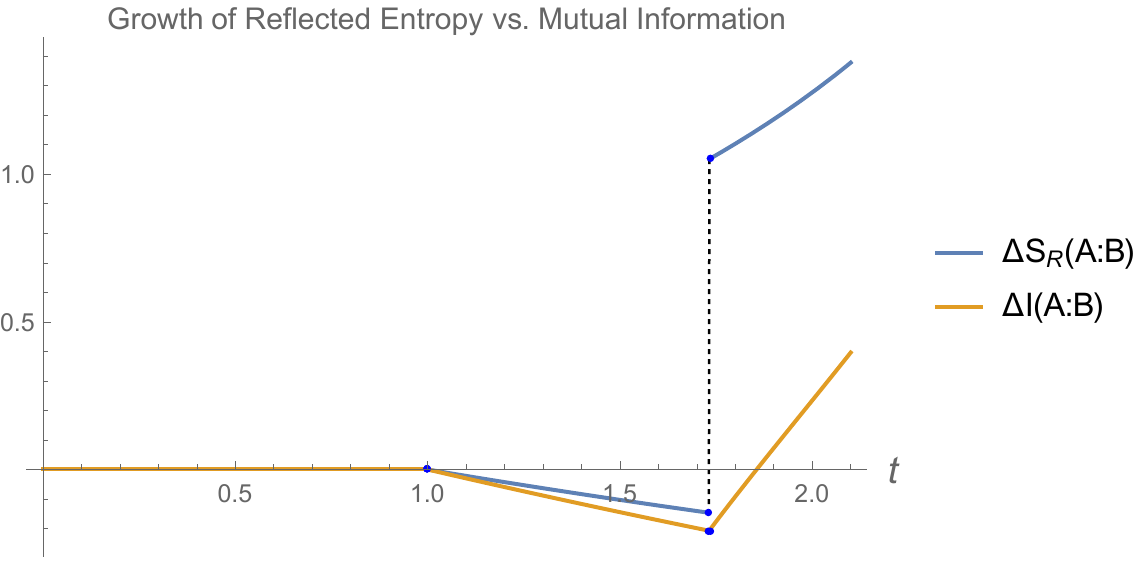}
  }
 \end{center}
 \caption{Reflected entropy (blue) and mutual information (yellow) for a state locally quenched outside two intervals. Here we have set $(u_1,v_1,u_2,v_2)=(-10,-3,1,20)$, $\e=10^{-3}$, $\g=2$ and we remove the prefactor $\fr{c}{6}$.
Each blue dot shows a transition of itself or its first derivative.}
 \label{fig:plot1}
\end{figure}

In this setup, we can summarize our results as follows (see also FIG. \ref{fig:plot1}):
For $t<-v_1$ or $-u_1<t$, we have
\begin{equation}\label{eq:main}
S_R (A:B)[O]=\dfrac{c}{6}\log\fr{1+\s{x}}{1- \s{x}}+\dfrac{c}{6}\log\fr{1+\s{\bar{x}}}{1- \s{\bar{x}}}, 
\end{equation}
where $(x,\bar{x})$ is given by
\begin{align}
&(x,\bar{x})=\nonumber\\
&\left\{\begin{array}{ll}
\left(\fr{(v_1-u_1)(v_2-u_2)}{(u_2-u_1)(v_2-v_1)} ,\fr{(v_1-u_1)(v_2-u_2)}{(u_2-u_1)(v_2-v_1)} \right),
			& (t<u_2) ,\\
   \left(\fr{(v_1-u_1)(v_2-t)}{(t-u_1)(v_2-v_1)}\fr{(v_1-u_1)(v_2-u_2)}{(u_2-u_1)(v_2-v_1)}\right) ,  & (u_2<t<\s{-v_1 u_2}) ,\\
    \left(\fr{(v_1-u_1)(v_2+t)}{(u_1+t)(v_1-v_2)},\fr{(v_1-u_1)(v_2-u_2)}{(u_2-u_1)(v_2-v_1)}\right), & (\s{-v_1 u_2}<t<-v_1) ,\\ 
		\left(\fr{(v_1-u_1)(t+u_2)}{(u_2-u_1)(t+v_1)}, \fr{(v_1-u_1)(v_2-u_2)}{(u_2-u_1)(v_2-v_1)}\right), & (-u_1<t<\s{-u_1 v_2}) ,\\ 
		\left(\fr{(v_1-u_1)(t-u_2)}{(u_2-u_1)(t-v_1)},
		\fr{(v_1-u_1)(v_2-u_2)}{(u_2-u_1)(v_2-v_1)}\right),  & (\s{-u_1 v_2}<t<v_2) ,\\ 
     \left(\fr{(v_1-u_1)(v_2-u_2)}{(u_2-u_1)(v_2-v_1)} ,\fr{(v_1-u_1)(v_2-u_2)}{(u_2-u_1)(v_2-v_1)} \right),& (v_2<t).
    \end{array}
  \right.
\end{align}
On the other hand, for $-v_1<t<-u_1$, we have obtained
\begin{align}
S_R (A:B)[O]&=\fr{c}{6} \log \left[ \fr{4(t+u_1)(t+u_2)(t+v_1)(t+v_2)}{\e^2(u_2-v_1)(u_1-v_2)}\right. \nonumber\\
&\hspace{-1cm}\left.\pa{\fr{\sinh \pi \bar{\g}}{\bar{\g}}}^2\right]+\fr{c}{6}\log\fr{1+\s{\fr{(v_1-u_1)(v_2-u_2)}{(u_2-u_1)(v_2-v_1)}    }}{1- \s{\fr{(v_1-u_1)(v_2-u_2)}{(u_2-u_1)(v_2-v_1)}  }  }.
\end{align}
Here we defined $\g=\s{\fr{24}{c}h_O-1}$ and $\bar{\g}=\s{\fr{24}{c}\bar{h}_O-1}$, where $h_O$ ($\bar{h}_O$) are the conformal dimension of the operator $O$. 
The above results are perfectly consistent with the EWCS in the falling particle geometry\cite{Kusuki:2019e}. 
In what follows, first we discuss which type of correlation is dominant in each time region. Second, we compare the results for holographic CFTs with ones for RCFTs. Finally we comment on the origin and importance of classical correlation for chaotic systems. 

The time region $t\notin [-v_1,-u_1] $ includes neither UV cutoff $\epsilon$ nor the information of local operators. This implies that we have only classical correlations between $A$ and $B$. 
In fact, the RE is more sensitive to classical correlations than the mutual information\footnote{The RE is always grater than the mutual information. Since our analysis in CFT is consistent with the entanglement wedge cross section, we can also relate the above discussion to original conjecture, the holographic entanglement of purification (EoP) $E_P (A:B)$\cite{Takayanagi2018a, Nguyen2018}. In particular, the EoP is more sensitive to the classical correlation than the RE, thus the importance of classical correlation becomes more remarkable. (For example, we have the lower bound of EoP for any states $E_P (A:B)\geq I(A:B)/2$, whereas we have the stronger lower bound for separable states $E_P (A:B)\geq I(A:B)$\cite{Terhal2002}. )}. 
Furthermore, this is consistent with the growth of RE and mutual information. 
In the lower two plots in FIG. \ref{fig:plot1}, we show the difference between the local quench state and the vacuum state, 
\begin{align}
\D S_R(A:B)&=S_R(A:B)[O] - S_R(A:B)[\bb{I}],\\
\D I(A:B)&=I(A:B)[O] - I(A:B)[\bb{I}],
\end{align}
which measure a growth of correlations after a local quench. 
We find the following inequalities for the mutual information and RE,
\begin{equation}\label{eq:DSR}
\begin{aligned}
\left\{
    \begin{array}{ll}
    \D S_R(A:B)  \geq \D I(A:B)  ,& \text{if } t\notin [-v_1,-u_1]  ,\\ \\
     \D S_R(A:B) \leq \D I(A:B)   ,& \text{if }  t\in [-v_1,-u_1] .\\
    \end{array}
  \right.\\
\end{aligned}
\end{equation}
On the other hand, the time region $t\in [-v_1,-u_1]$ mainly consists of quantum correlations. This can be understood from the well-known fact that the mutual information for holographic CFTs measure mostly the quantum correlation\cite{Hayden2013}. Indeed, \eqref{eq:DSR} shows that growth of the mutual information is greater than one of the RE in this region .  

Finally we discuss the origin and importance of classical correlations by comparing with the results for RCFTs. It turns out that the same analysis for RCFTs can be understood from the quasi-particle picture (left-right propagation of ``EPR pair'') just as the same as the mutual information in RCFTs. Importantly, we have $\D S_R(A:B) = \D I(A:B) =0$ for $t\notin [-v_1,-u_1] $, whereas we have some classical correlations for chaotic theories, at least holographic ones which is (in some sense) the most chaotic systems. This classical correlation basically comes from the process for creating our local operator quench state. In the strict sense, we cannot create our local excitation via the local operation at $t=0$. This should give rise to additional correlations. Furthermore, our chaotic systems, where we have large amounts of degrees of freedom and strong interactions, enhance any correlations significantly. 
In summary, the existence of classical correlations directly suggest the chaotic nature of a given system in the light of RE and the mutual information. Since the former is more sensitive to the classical correlations, we can expect the RE is a more sensitive criteria for whether a given system is chaotic or not.

\section{Heavy state and subsystem ETH}
We consider a CFT on a circle with length $L$. Then, the RE for a heavy primary state can be obtained from
\begin{align}\label{eq:Renyi2}
Z_{n,m}=\langle O^{\otimes mn}| \sigma_{g_A}(u_1)\sigma_{g_A^{-1}}(v_1)\sigma_{g_B}(u_2) \sigma_{g_B^{-1}}(v_2)| O^{\otimes mn}\rangle.
\end{align}
Here, this correlator is defined on a cylinder. This can be mapped to the plane $(z,\bar{z})$ by
\begin{equation}
z=\ex{\fr{2 \pi i w}{L}}, \ \ \ \ \bar{z}=\ex{-  \fr{2\pi iw}{L}}.
\end{equation}

For a sufficiently large subsystem, under the large-$c$ limit, we have obtained
\begin{align}\label{eq:REinH}
S_R(A:B)&=
\fr{c}{6} \log \pa{\coth\fr{ \pi \g   (u_2-v_1)}{2L}} \nonumber\\
&+ \fr{c}{6} \log \pa{\coth\fr{ \pi \bar{\g}   (v_2-u_1)}{2L}}.
\end{align}

\begin{figure}[t]
 \begin{center} \resizebox{80mm}{!}{
  \includegraphics{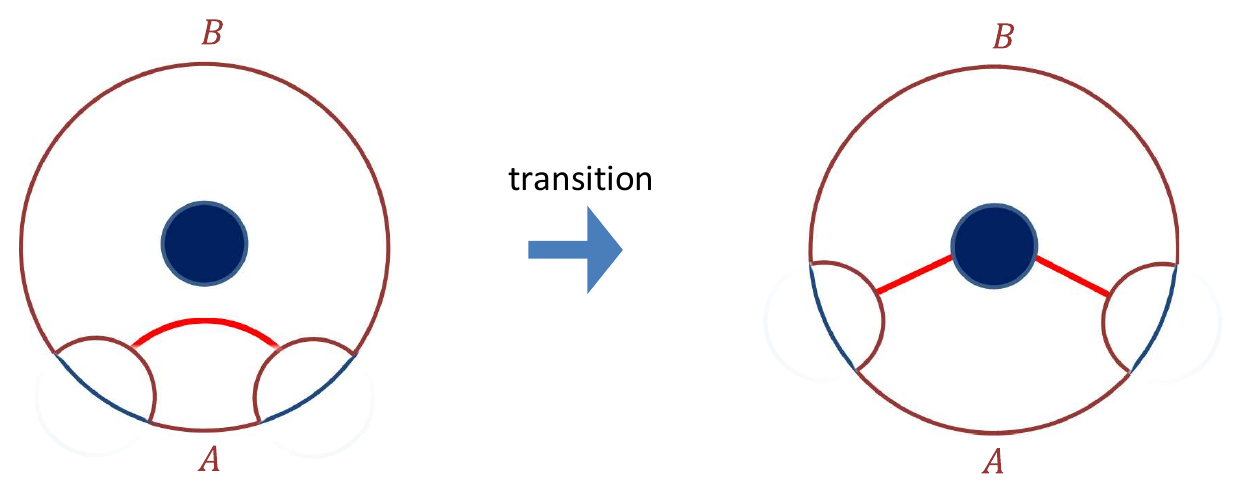}}
 \end{center}\vspace{0cm}
 \caption{The non-trivial entanglement wedge cross section in the BTZ black hole has two phases. 
We can also see this transition from the evaluation of pure state in \eqref{eq:Renyi2}. An important point is that if we evaluate the usual EE for thermal state, the similar phase as right panel never appear, whereas the pure state does. The existence of second phase in our pure state result means that after this transition, we cannot approximate pure state as thermal one.}
 \label{fig:transition}
\end{figure}

This result perfectly matches the entanglement wedge cross section in the BTZ metric \cite{Takayanagi2018a}, namely the cross section described in the right panel of the FIG. \ref{fig:transition}. 
It means that the thermalization in the large $c$ limit \cite{Fitzpatrick2015, Lashkari2018, Hikida2018, Romero-Bermudez2018, Brehm2018} can also be found in the RE. (For a sufficiently small subsystem, we can also obtain one for left panel of the same figure. )

On the other hand, we have shown that the surface ends at the horizon of the black hole. This can be explained by considering the horizon as an end of the world brane \cite{Hartman2013, Almheiri2018, Cooper2018}. In this case, the surface can end at the horizon even if we consider a pure state black hole. We stress that this is also the case for EE in a heavy primary state because the RE (\ref{eq:REinH}) should reproduce the double of the EE in the pure state limit\footnote{Note that the pure state limit of the \eqref{eq:REinH} does not match the result in \cite{Asplund2015}. This is because their derivation implicitly assumes that the change of the dominant channel (i.e., the transition shown in FIG. \ref{fig:transition}) does not happen. However, the result under such an assumption contradicts the pure state limit, and basically there is no reason to remove the possibility of the transition even in the EE.}.
Notice that this phase transition does not appear in the holographic EE for BTZ black hole, namely the true thermal state. Therefore, the transition point tells us how large subsystem (the reduced density matrix) can pretend the thermal system. Such imitation is called subsystem eigenstate thermalization hypothesis (subsystem ETH)\cite{Garrison:2015lva,Dymarsky:2016ntg}. Previously, we expected this transition point would happen at the half of the subsystem, whereas our result did prove this is actually not the case for heavy primay states. 
Now the transition length for single-interval EE turns out to be $(2\pi/\gamma)\log(1+\sqrt{2})<\pi$ where we took $L=2\pi$ and $\gamma=\bar{\gamma}$, for simplicity. We can easily derive it from the pure state limit of the two phases. In particular, under the high-energy limit $2\pi/\gamma\rightarrow0$, this transition does happen quickly. Such distinguishability can be also seen from the Holevo information\cite{Bao:2017guc,Michel:2018yta}, for example.

\if(

\section{RCFT}
It is very interesting to compare our result to the dynamics of the RE in other CFTs, especially RCFTs. 

If we consider the setup ( $0<\e\ll u_2<-v_1<-u_1<v_2$ and $O$ is acted on $x=0$ at $t=0$.) for example, we obtain
\begin{equation}\label{eq:DSRR}
\begin{aligned}
\D S_R(A:B)[O]&=\left\{
    \begin{array}{ll}
  0 
		,   & \text{if } t<-v_1 ,\\ \\
  2\log d_O 
		,   & \text{if } -v_1<t<-u_1,\\ \\
  0
		,   & \text{if } -u_1<t ,
    \end{array}
  \right.\\
\end{aligned}
\end{equation}
where $d_O$ is a constant, so-called quantum dimension, which is re-expressed in terms of the modular S matrix as \cite{Numasawa2016, He2014}
\begin{equation}
d_O=\fr{S_{0O}}{S_{00}}.
\end{equation}
One can find two significant differences from FIG.\ref{fig:RCFTplot},
\begin{itemize}
\item The small effect in  $ t\in [u_2, -v_1] \cup [-u_1, v_2]$ does not appear in RCFTs, unlike the holographic CFT.

\item The holographic CFT shows the logarithmic growth in $ t\in [-v_1, -u_1]$, on the other hand, the growth of RCFT approaches a finite constant.
\end{itemize}

It would be interesting to note that this growth pattern (\ref{eq:DSRR}) is exactly the same as that of the mutual information, which is quite natural for RCFTs because the quasi particle picture can be applied in any time region.

\begin{figure}[t]
 \begin{center}
 \resizebox{60mm}{!}{
  \includegraphics{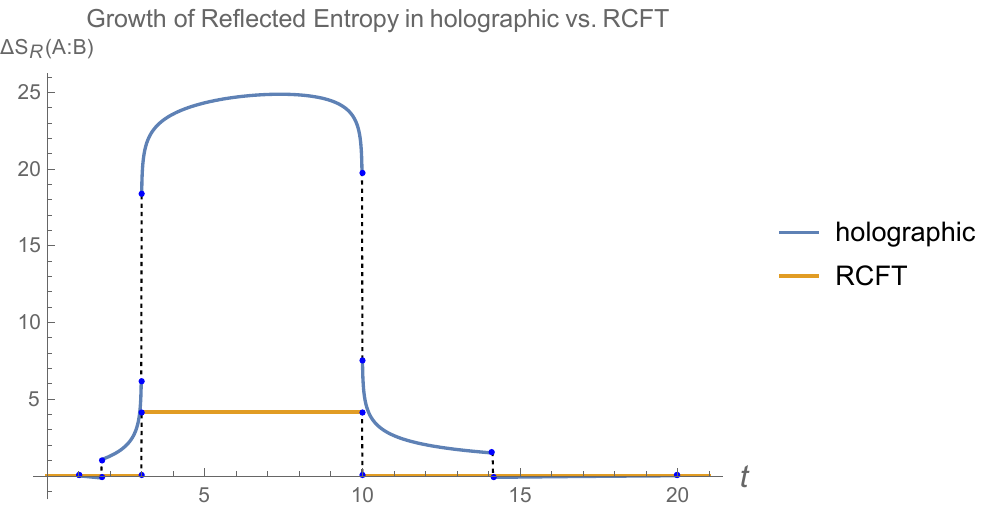}
  }
 \end{center}
 \caption{ The growth of reflected entropy in holographic CFT (blue) and Ising model (yellow). 
$\D S_R$ means the difference between the excited state and the vacuum state.
Here $(u_1,v_1,u_2,v_2)=(-20,-1,3,10)$, $\e=10^{-3}$
and we divide them by $\fr{c}{6}$. We choose $\g=2$ in holographic CFT and $O=\sigma$ in Ising model.
Each blue dot shows a transition of itself or its first derivative.
}
 \label{fig:RCFTplot}
\end{figure}

)\fi

\section{Discussion}

One can also reproduce the above results from the odd entanglement entropy\cite{Tamaoka2019} by replacing $S_R(A:B)/2$ with the odd EE minus von-Neumann entropy for the above results (In fact, this is also the case for RCFTs). 
This coincidence can happen because we are considering large $c$ limit and/or Regge limit which give us quite universal consequences. In more general parameter regimes, these two quantities should behave differently. It is very interesting to study further such regimes.
\if(which is defined by
\begin{equation}\label{eq:OEE}
S_o(A:B)\equiv\lim_{n_o\to1} \fr{1}{1-n_o}\BR{\tr \pa{\rho_{AB}^{T_B}}^{n_o}-1},
\end{equation}
where $\rho_{AB}$ is a reduced density matrix for subsystems $A$ and $B$, obtained by tracing out its complement. The limit $n_o \to 1$ is the analytic continuation of an odd integer and $T_B$ is the partial transposition with respect to the subsystem $B$. The odd EE for holographic CFT is expected to have the following relation,
\begin{equation}\label{eq:OEEdef}
S_o(A:B)-S(AB)=E_W(A:B).
\end{equation}
Indeed, as like the original proposal in \cite{Tamaoka2019}, one can replace $\frac{1}{2}S_R(A:B)$ with the OEE minus von-Neumann entropy for the above results (In fact, this is also the case for RCFTs). 
This coincidence can happen because we are considering large $c$ limit and/or Regge limit which give us quite universal consequences. In more general parameter regimes, these two quantities should behave differently. It is very interesting to study further such regimes. )\fi

We gave a counterexample of the subsystem ETH, a heavy primary state in two-dimensional holographic CFTs. One possibility might be that such state is not typical. Since this is still counterintuitive, we should deepen our understanding of this fact as this state has been often discussed as an explicit example of typical state in literature. 


Finally, there are several interesting future directions which can be accomplished in a similar manner. For example, it would be interesting to understand a relation to negativity \cite{Kudler-Flam2019}, to study dynamics in other irrational CFTs \cite{Caputa2017,Caputa2017a}, to investigate information spreading by using the RE \cite{Asplund2015a}, and evaluate the Renyi RE, in particular, its replica transition \cite{Kusuki2018b,Kusuki2018c,Kusuki2019}.
\section*{Acknowledgments}
We thank Souvik Dutta, Thomas Hartman, Jonah Kudler-Flam, Masamichi Miyaji, Masahiro Nozaki, Tokiro Numasawa, Tadashi Takayanagi and Koji Umemoto for fruitful discussions and comments. YK is supported by the JSPS fellowship. KT is supported by JSPS Grant-in-Aid for Scientific Research (A) No.16H02182 and Simons Foundation through the ``It from Qubit'' collaboration. We are very grateful to ``Quantum Information and String Theory 2019'' and ``Strings 2019'' where the finial part of this work has been completed. 

\bibliography{multi}

\if(
\pagebreak
\widetext
\begin{center}
\textbf{\large Supplemental Materials}
\end{center}
\setcounter{equation}{0}
\setcounter{figure}{0}
\setcounter{table}{0}
\setcounter{page}{1}
\makeatletter
\renewcommand{\theequation}{S\arabic{equation}}
\renewcommand{\thefigure}{S\arabic{figure}}
\renewcommand{\bibnumfmt}[1]{[S#1]}
\renewcommand{\citenumfont}[1]{S#1}

\section{Section 1}
Copy and paste your Supplemental Materials text here \cite{S_RefA}, blah, blah, blah, blah, blah, blah, ...
\be
aaa
\ee

)\fi

\end{document}